# Possible Stimulated Emission of Entangled Rhodium Mössbauer Gammas

Yao Cheng, Zhongming Wang

Department of Engineering Physics, Tsinghua University, Beijing 100084

Email: yao@tsinghua.edu.cn

**Abstract**

Observation of possible stimulated emission of Mössbauer gamma is reported by liquid-nitrogen quenching of rhodium sample from room temperature to 77K in the time-resolved Mössbauer spectroscopy. Recently, we have demonstrated the anomalous emission of three entangled gammas of the E3 Mössbauer transition generated by bremsstrahlung irradiation. In this work, we further report the high-speed decay of excited state. We conjecture that cooling shrinkage, gravitational redshift and crystal lattice collimate entangled gammas in a linear cavity. This opens up a new approach towards gamma lasing, if the stimulated emission occurs at this obtained low excitation density.

To create solid-state graser (gamma-ray laser) [1], the following necessary prerequisites have to be satisfied; 1: cavity, 2: active medium 3: sufficient gain. In this letter, we demonstrate that these necessary conditions may be fulfilled by liquid-nitrogen quenching in the time-resolved Mössbauer spectroscopy. Linear cavity of the coherent nuclear emission is induced by quenching shrinkage with gravitational redshift [2]. In the course of this study [3-5], we generate the long-lived E3 Mössbauer transition of rhodium by bremsstrahlung irradiation. Mössbauer gammas are partially entangled in the Borrmann channel with suppressed photo-electric attenuation [6] and suppressed Doppler shift [4]. Resonant nuclear scattering thus dominates the photon transport in channel. The suppression of K internal conversion as a function of the excitation density was already attributed to stimulated emission in the early stage of these series of works [5]. The low dimensionality of photon propagation realized recently puts forward this speculation. As a matter of fact, the achievable excitation density obtained from bremsstrahlung irradiation is far from population inversion, graser can be realized only by lasing without inversion.

The frequency shift induced by temperature and gravitation is

$$\frac{\nu}{\nu_0} = \frac{1+\beta\cos\theta}{\sqrt{1-\beta^2}} - \frac{g\Delta z}{c^2} - \frac{C_L \Delta T}{2Mc^2}, \qquad (1)$$

where the frequency ν at moving frame deviates from the natural Mössbauer frequency of $\nu_0$. The three terms are attributed to the relativistic drift with the velocity β of quenching shrinkage, the gravitation acceleration g, and the thermal motion of nuclei [7] respectively. θ is the angle included between the linear cavity and β. Δz is the relative height and c is the light speed in vacuum. ΔT is the temperature deviation, M is the gram atomic weight of rhodium and $C_L$ is the specific heat of the lattice [7]. Formula (1) based on the conventional understanding of the Mössbauer effect helps us here to describe the linear cavity confined by quenching shrinkage. Since the $10^{-19}$ eV natural linewidth of rhodium Mössbauer emission corresponds to the Doppler velocity of fm/s, the first term in (1) is dominant. Quantitative estimations of the second order Doppler shift in the first term and the third term in (1) give $\beta=10^{-13}$ and $\Delta T=10^{-9}$ K for one linewidth respectively. If this estimation is overall valid, one should not be able to observe any resonance under this experimental environment [3-5]. Consequently, Doppler shift is suppressed for the entangled photon transport in the Borrmann channel.

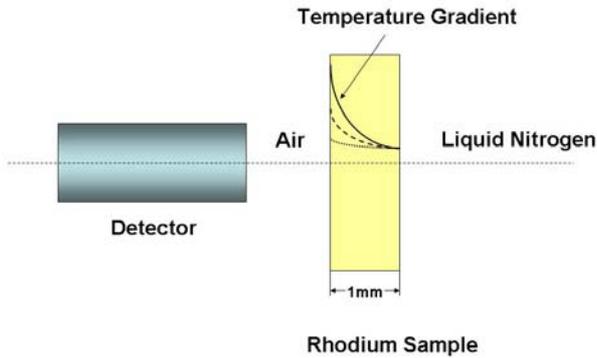

FIG. 1: Experimental setup. Typical evolution of the temperature gradient cross the sample is produced by the single–side quenching.

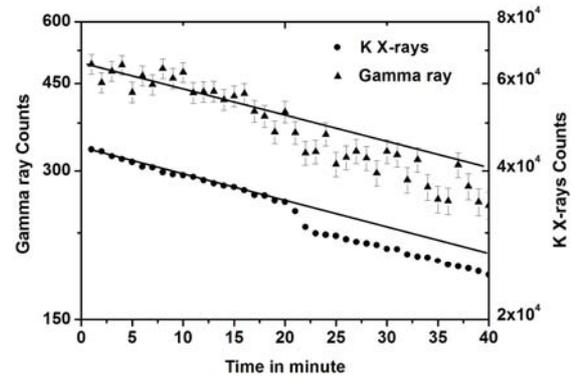

FIG. 3: Blown-up of the first forty minutes of Figure 1. The quenching of liquid nitrogen occurs at the twenty-first point.

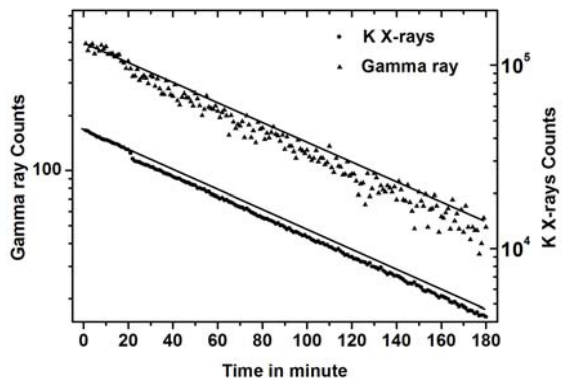

FIG. 2: Time evolution of gamma and K X-rays emitted from the long-lived Mössbauer transition. Temperature was lowered by liquid nitrogen for 24 min after 20 min measurement in room temperature. The K counts are collected from 35 channels between 20 keV and 23.5 keV. Gamma is collected from 16 channels centered at 39.8 keV. Every collection period is one minute. The right ordinate is the K counts and the left ordinate is the gamma counts.

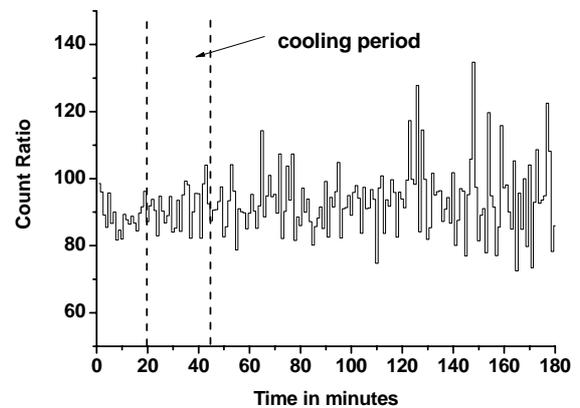

FIG. 4: Time evolution of the internal conversion K/γ. Each data is counted in one minute and 40 channels for K and 16 channels for γ. The 2% pile-up of Kα at the right shoulder of gamma has been removed from the denominator.

The experimental conditions such as detector, excitation procedure, and sample cooling are exactly the same as detailed in [3], except that the cooling is now real single-side quenching without overflowing the liquid nitrogen to the other side. The total initial count rate is around 700 cps, which is almost the same count rate as obtained in [3]. We kept the sample in room temperature of 18°C for twenty minutes and then poured in liquid nitrogen for 24 minutes. The sample temperature recovered slowly back to room temperature one hour later as described in [3]. Our sample is a heat source after irradiation. Temperature gradient existed in the direction from the detector side to the liquid-nitrogen contact as illustrated in Fig. 1. Its relaxation time constant shall be in the order of 10 seconds, estimated from the heat capacity and heat conductance of the rhodium sample. Time evolutions of K X-rays and γ are illustrated in Fig. 2. A significant loss (>10%) of the nuclear excited states during the first three minutes of quenching is observed (Fig. 3), whereas the referential emissions such as $^{195m}$Pt and impurity scatterings [3] stay unchanged during the corresponding period of loss. The lost count rate did not recover till the end of 3 hr measurement, when temperature approached room temperature. On the contrary, the speed-up decay in [3] recovered immediately at the moment of quenching stop. Reduction of internal conversion coefficient was treated to be one of the stimulated emission features [3]. However, the internal conversion of Fig. 4 obtained from the sample surface will not demonstrate the K suppression, if stimulated emission occurs in the quenching side. Count evolutions in the same energy regions as reported in [3] are illustrated in figures 5 and 6. Similar count enhancement is observed during the cooling period. The major difference between these two measurements reported here and in [3] is the time delay of count enhancement in Fig. 6.

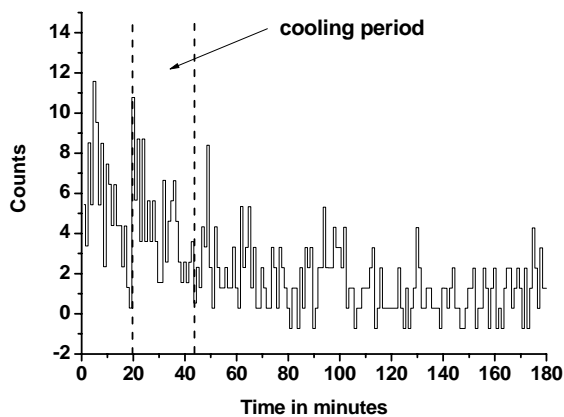

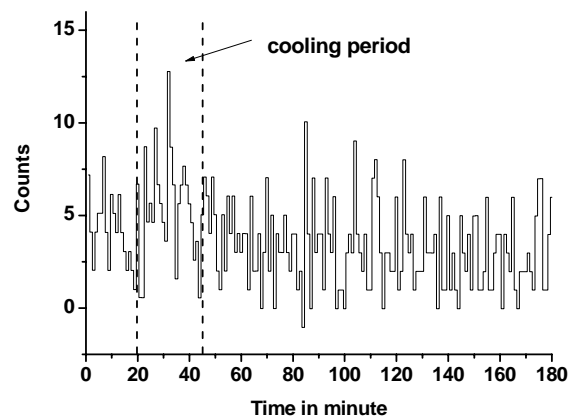

FIG. 5: Time evolution of the K-γ and triple k pile-ups between 46 keV and 63 keV. Every collection period is one minute. The average background has been removed.

FIG. 6: Time evolution of the tri-gamma scatterings between 63 keV and 90 keV. Every collection period is one minute. The average background is subtracted.

Superradiance channels open in the particular direction at Bragg. This will not affect the effective total decay rate, which is averaged over all possible superradiant and subradiant modes [8]. The immediate recovery of decay rate at the sample surface after quenching reported in [3] indicates the closing of the superradiance channel. This was recognized as being the collective anomalous-emission effect to generate tri-gamma [3]. The Hannon-Trammell anomalous emission is due to the constructive interference of nuclear multipoles emitted into the Borrmann channel [8]. Lowering the sample temperature will enhance this effect but not the stimulated emission because of the negative gain. Population inversion is thus required to provide gamma lasing [1]. The excitation density obtained from the reported bremsstrahlung irradiation is about $10^{12}$ cm$^{-3}$, which is far from the required value for amplification. An open mechanism of inversionless amplification [9] or Raman lasing is necessary to generate the stimulated emission.

Based on the observed loss after quenching in this letter, we surmise that simulated tri-gamma emission of the Mössbauer transition occurred, through matching of several factors to maintain sufficient gain. These factors are the low dimensionality, low temperature, slow temperature changing rate and low spatial temperature gradient. Therefore, the stimulated emission shall take place first at the quenching side and then propagates to the opposite side of the sample. Simulated emission occurs in the linear cavity, which is along the long axis perpendicular to the detecting direction due to the temperature gradient. No stimulated gamma is thus directly observed.

We are indebted to Long Wei and Jin Li at Institute of High-Energy Physics, Chinese Academy of Sciences for there kindness to lend us the level HPGe detector. We thank Hong-fei Wang for help on manuscript preparation, Yexi He for fruitful discussions, Bing Xia for data collection, Mingda Li for the proofreading and Yuzheng Lin for his accelerator team.